# Exchange-induced suppression of superconductivity in a nano-skyrmion lattice - superconductor hybrid

Dongfei Wang, [1†] Wenbin Li, [2†✉] Eric Mascot,[2,3†] Roland Wiesendanger[2✉]

[1]School of Physics; Centre for Quantum Physics, Key Laboratory of Advanced Optoelectronic Quantum Architecture and Measurement (MOE); State Key Laboratory of Chips and Systems for Advanced Light Field Display; Beijing Institute of Technology, 100081 Beijing, China

[2]Department of Physics, University of Hamburg, D-20355 Hamburg, Germany.

[3]School of Physics, University of Melbourne, 3010 Victoria, Australia.

[†]These authors contributed equally: Dongfei Wang, Wenbin Li, Eric Mascot.

[✉]Email: li@spm.ac.cn; wiesendanger@physnet.uni-hamburg.de

## Abstract

Engineered magnet-superconductor hybrids have recently been identified as promising platforms for the investigation of topological superconductivity. Beyond ferro- and antiferromagnetic structures, coupling non-collinear spin textures, such as skyrmion lattices, to superconductors offers an exciting route for creating and manipulating unconventional superconducting states. In this work, by preparing monolayer Fe on Ir(111) thin films grown epitaxially on a Nb(110) surface, we realize a hybrid system of a nano-skyrmion lattice proximitized to a superconducting substrate. Scanning tunneling spectroscopy shows that superconductivity becomes suppressed by the Fe nano-skyrmion lattice, with both the superconducting gap and coherence peaks disappearing. Tight-binding calculations reveal that with increasing exchange coupling, the gap is progressively filled up and eventually superconductivity gets quenched. These results reveal microscopic constraints for designing topologically non-trivial states based on magnet-superconducting heterostructures.

## Introduction

Magnetic nanostructures coupled to conventional s-wave superconductors have emerged as a powerful platform to engineer *field-free* topological superconductivity and Majorana modes [1,2]. The basic idea is to lower the effective spin singlet pairing and increase triplet pairing using specially designed magnet-superconductor heterostructures. Under suitable combination of time-reversal, particle-hole as well as chiral symmetry, the magnet-superconductor hybrid can enter a topological superconducting phase [3]. Following this principle, topological superconductivity and signatures of Majorana modes have been pursued based on several engineered platforms, including magnetic atom chains on superconductors [4–10] and single-layer 2D magnets on top of a superconducting substrate, such as Fe/Re(0001) [11], $CrBr_3/NbSe_2$ [12] and Mn/Nb(110) [13]. Within this broader framework, magnetic skyrmions provide an attractive form of magnetism because their nanoscale non-collinear spin textures are stable and topologically protected [14,15]. Moreover, when skyrmions interact with a superconductor, a topological superconducting phase may emerge [16–18], and Majorana modes will appear at the center of isolated skyrmions [19–22] or at the boundary of a skyrmion lattice [17,23–25].

As demonstrated previously, skyrmions can be stabilized, switched, or moved by external stimulations, including local spin current injection [26], local electric fields [27], magnetic fields [28], and electrical currents [29]. Thus, controllable Majorana bound state movement in a skyrmion-superconductor hybrid system becomes feasible by manipulating the corresponding skyrmion [30]. However, the preparation of a skyrmion-superconductor hybrid is experimentally challenging. Many experimentally established skyrmion systems are realized in magnetic films of considerable thickness or multilayer stacks. In those material systems, the magnetic volume and resulting dipolar stray fields can strongly disturb a nearby superconductor [28,31–33] by inducing vortices or locally suppressing superconductivity.

Recent experiments have demonstrated the synthesis of skyrmion-superconductor heterostructures [34,35] and realized the skyrmion-Pearl vortex interaction [36]. These studies focused on skyrmion-vortex coupling and vortex manipulation, but they did not directly reveal the exchange-coupling mechanism required for skyrmion-induced topological superconductivity and Majorana modes. Single-atomic-layer spin textures provide a more suitable platform for studying the direct interaction between a skyrmion lattice and a superconducting substrate. Unfortunately, atomic-scale skyrmion lattices in monolayer films remain rare, with Fe monolayers on Ir(111) being the most established example [37,38]. Isolated zero-field skyrmions or low-field induced skyrmions have been observed in Pd/Fe bilayers on Ir(111) [39] or Co monolayers on Ru(0001) [40]. However, the coupling between skyrmions and the superconductor's electronic states and the question whether such coupling can support topological superconductivity remained largely unexplored.

In this work, we realize a direct-coupled skyrmion-superconductor hybrid by growing monolayer Fe, exhibiting a nano-skyrmion lattice, on a proximitized superconducting Ir(111) thin film. The Fe/Ir(111) nano-skyrmion lattice has been chosen because of its nearly compensated spin texture, which helps minimizing stray-field effects. Nb(110) is used as the superconducting substrate because of its highest transition temperature among all elemental superconductors, which helps inducing a sizable proximity gap in the Ir(111) thin film. Experimentally, we first grew an extended flat Ir(111) thin film on the Nb(110) substrate employing atomic layer deposition techniques. Subsequently, Fe islands with single-layer thickness and tens of nanometers in size were prepared on the as-grown Ir(111) thin film by e-beam deposition. Using spin-polarized scanning tunneling microscopy (SP-STM), we identified the *field-free* atomic-scale skyrmion lattice on the two-dimensional Fe islands. Using low-temperature scanning tunneling spectroscopy (STS), we measured and compared the differential tunneling conductance (dI/dV), reflecting the low-energy local density of states (LDOS), on the Nb(110) substrate, the Ir(111) thin film, the Fe nano-skyrmion lattice, and the exposed Ir(111) surface near the skyrmion lattice. Our results show clear superconducting gaps on both the Nb(110) substrate and the proximitized Ir(111) thin film. In contrast, a gap closing effect was observed on the Fe islands revealing the nano-skyrmion lattice. By comparing the spatially resolved tunneling spectra with tight-binding calculations, we identify the strong direct exchange coupling as the main factor responsible for the quenching of superconductivity on the Fe islands.

**Experimental results and discussion**

Figure 1(a) shows the almost oxygen-free surface of Nb(110) prepared using the same cleaning method as reported previously [41]. After e-beam deposition of Ir, the surface is fully covered by a flat Ir thin film with some nano-scale islands of the next Ir layer (Fig. 1(b)). The Ir thin film follows the step-and-terrace structure of the Nb(110) substrate and has a thickness of about 5 nm, which is thin enough for proximity-induced superconductivity. Atomic-resolution STM images taken on the flat surface regions show a triangular lattice with a lattice constant of 0.27 nm (Fig. 1(c)), which is consistent with the Ir(111) lattice constant [42–44]. The low-energy dI/dV spectra taken on the Ir(111) thin film is shown in Fig. 1(d). The Dyne's fitting reveals a proximity-induced superconducting gap of 1.16 meV. By comparing the gap value of 1.53 meV as measured on bare Nb(110) [41], we find that the gap value of the Ir(111) thin film is reduced to 76% of the Nb value, but remains large enough for studying the interaction between the nano-skyrmion lattice and proximitized superconductivity.

Next, Fe was deposited onto the Ir(111) thin film surface at room temperature. The STM topography of the sample surface after the Fe deposition is shown in Fig. 2(a), where island-like structures are observed on the Ir surface. Since the topographic contrast alone is insufficient to unambiguously distinguish Fe-covered regions from the exposed Ir surface, we additionally revealed the local electronic contrast variations by dI/dV mapping at -800 mV sample bias. As shown in Fig. 2(b), the Fe-covered regions exhibit a darker (lower) dI/dV contrast compared to the surrounding Ir surface, allowing the triangular-shaped Fe islands as well as the Fe patches nucleated along the step edges of the Ir film to be clearly identified. We then examined whether the Fe islands exhibit a similar characteristic magnetic structure as Fe monolayers prepared on Ir(111) single crystals using SP-STM. At low bias voltages, such as 5 mV and 20 mV, the spin-resolved dI/dV and current images show nearly square patterns with a periodicity of about 1 nm on the Fe islands (Fig. 2(c-f)). The observed periodicity and symmetry of these patterns are consistent with the presence of a nano-skyrmion lattice as previously reported for single-layer Fe on Ir(111) single crystals [37], demonstrating that the atomic-scale skyrmion lattice is formed in the monolayer Fe patches on Ir(111) thin films being in direct contact with the superconducting Nb(110) substrate.

We next performed low-temperature STS studies across a nano-skyrmion Fe island and the surrounding Ir film surface to probe the spatial evolution of proximitized superconductivity. Figure 3(a) shows a spin-polarized STM image of an Fe island on the Ir(111) thin film, where the nano-skyrmion lattice can be clearly recognized. Spatially resolved dI/dV spectra were then acquired along the green arrow starting from the uncovered Ir film region and going across the nano-skyrmion lattice (Fig. 3(b)). Along the uncovered Ir region, from position A to B, the dI/dV spectra still show a superconducting gap feature, although the gap is partially filled compared with that measured on the clean Ir(111) film before Fe deposition. In contrast, after the STM tip is located above the Fe nano-skyrmion lattice, from position C to the end of the line, the superconducting gap is fully filled, and the shape of the dI/dV spectrum remains nearly unchanged across the Fe island. This behavior is further confirmed by representative tunneling spectra obtained at the positions A-D marked in Fig. 3(a), which correspond to the uncovered Ir surface, the Ir region near the Fe island edge, the Fe island edge, and the center of the Fe island, respectively (Fig. 3(c)). Compared with the reference spectrum measured on the clean Ir(111) thin film before Fe deposition, these spectra show that superconductivity is completely quenched on the Fe island, while the proximitized superconducting gap on the uncovered Ir surface is partially suppressed. Numerous other spectra acquired on various Fe islands and uncovered Ir regions show the same behavior, confirming the complete gap quenching on Fe islands and partial gap suppression on the surrounding Ir film surface (Fig. S1).

In addition to the overall gap suppression, an enhanced zero-bias signal is observed near the Fe island edge in the line spectra (Fig. 3(b)). This edge-localized in-gap enhancement is also visible in the two-dimensional dI/dV map (Fig. S2(c)). Such a boundary-related signal has been reported for topological non-trivial states originating from the coupling between a skyrmion lattice and a superconducting substrate. However, previous calculations [17] suggested that a full superconducting gap can be preserved inside the skyrmion lattice under the relevant symmetry protection. This, however, is in contrast to our experimental observation that the superconducting gap is fully quenched on the Fe island. Thus, the origin of the observed edge signal remains unclear and needs to be explored in more detail by theory.

To quantify the degree of superconducting gap suppression, we extracted a site-dependent LDOS ratio from the line spectra in Fig. 3(b). The ratio is derived from the dI/dV intensity at the Fermi level divided by the dI/dV intensity at the coherence peaks. For an ideal s-wave superconducting gap, this ratio should approach zero. With increasing gap filling and decreasing coherence-peak intensity, the ratio increases. When superconductivity is fully quenched, the ratio is expected to approach the value 1. As shown in Fig. 3(d), the ratio is about 0.6 at position A of the uncovered Ir(111) surface and increases as the measured position approaches the Fe island. Once the spectra are acquired on the Fe island, the ratio rapidly approaches 1, which quantitatively confirms the complete quenching of superconductivity on the Fe skyrmion lattice. Notably, an anomalous enhancement appears at the Fe island edge in the ratio profile, which is associated with the near-zero energy in-gap edge states enhancement discussed above.

We further derived a two-dimensional LDOS ratio map from the measured dI/dV maps (Fig. 3(e)) acquired at the Fermi level and at the coherence-peak energies (Fig. S2). The ratio map shows that superconductivity is nearly homogeneously quenched on the Fe island, with the ratio being close to 1. In the surrounding uncovered Ir regions, the ratio remains higher than 0.4, indicating that proximitized superconductivity is also substantially suppressed around the magnetic island. This residual suppression is likely related to the short distance between the measured Ir regions and nearby Fe islands. To further examine the distance dependence, we prepared another sample with lower Fe coverage, where individual Fe islands are more widely separated (Fig. S3(a)). Large-scale spatially resolved STS measurements on this sample show a clear recovery of the superconducting gap when the measurement position is about 10 nm away from the Fe island (Fig. S3(b)), thereby confirming the long-range, but distance-dependent suppression of proximitized superconductivity by the Fe islands.

**Theoretical calculations**

The complete quenching of superconductivity on a nano-skyrmion lattice is unexpected for an atomic magnetic layer on a superconducting surface. In previous studies of monolayer ferromagnetic or antiferromagnetic films on superconductors, superconducting gap features could often still be resolved on or near the magnetic layer [11,13,45–47]. In our present system, the nano-skyrmion lattice is also only one atomic layer thick and has a nearly compensated spin texture, which should strongly reduce stray-field-induced pair breaking. Therefore, the observed gap quenching is unlikely to be explained by a simple Zeeman effect from a magnetic stray field but instead points to a strong local exchange coupling between the Fe spin texture and the proximitized electronic states of the Ir(111) thin film. To capture this effect, we performed tight-binding Bogoliubov-de Gennes calculations for a periodic nano-skyrmion lattice (see Fig. S4) coupled to an s-wave superconducting substrate. The model Hamiltonian is written as:

$$H = \sum_{r,r',s} (-t_{r,r'} - \mu\delta_{r,r'})\, c_{rs}^{\dagger} c_{r's} + \Delta \sum_{r} \left( c_{r\uparrow}^{\dagger} c_{r\downarrow}^{\dagger} + \mathrm{H.c.} \right) + J \sum_{r,s,s'} \boldsymbol{S}_r \cdot c_{rs}^{\dagger} \boldsymbol{\sigma}_{ss'} c_{rs'}$$

$$+\, i\alpha_{\mathrm{R}} \sum_{r,r',s,s'} c_{rs}^{\dagger} \left[ \hat{\boldsymbol{z}} \cdot \left( \boldsymbol{\sigma}_{ss'} \times \hat{\boldsymbol{d}}_{rr'} \right) \right] c_{r's'} + \mathrm{H.c.}$$

where $c_{rs}^{\dagger}$ and $c_{rs}$ are the creation and annihilation operators for an electron with spin $s$ at lattice site $r$, where $s, s' = \uparrow, \downarrow$ denote spin indices. $t_{r,r'}$ is the hopping amplitude between sites $r$ and $r'$, and only nearest-neighbor hopping is considered, namely $t_{r,r'} = t$. $\mu$ is the chemical potential, and $\Delta$ is the onsite s-wave pairing potential induced by the superconducting substrate. $\mathbf{S}_r$ represents the local spin direction of the skyrmion lattice at site $r$, with the spin magnitude assumed to be uniform. $J$ denotes the onsite s-d exchange coupling between the local Fe spin and the itinerant electron spin. The last term describes Rashba spin-orbit coupling with strength $\alpha$, where $\boldsymbol{\sigma}$ is the vector of Pauli matrices, $\hat{\mathbf{z}}$ is the surface normal vector, and $\hat{\mathbf{d}}_{rr'}$ is the unit vector connecting neighboring sites $r$ and $r'$.

Following previous tight-binding BdG models for magnetic structures coupled to s-wave superconductors and effective Shiba-band descriptions [48,49], we used $t$ as the energy unit and set the induced pairing potential to $\Delta = 0.02\ t$. Since the microscopic parameters of the Fe/Ir(111)/Nb(110) interface are not known precisely, the calculation is not intended as a direct fit to the experimental spectra. Instead, we first used a representative set of parameters ($\mu = 0$ and $\alpha = 0.02\ t$) to examine how the superconducting DOS evolves with increasing onsite exchange coupling $J$. Figure 4(a-d) shows some representative calculated DOS near the Fermi energy for various values of $J$, with all other parameters fixed. At $J = 0$, a well-defined superconducting gap with pronounced coherence peaks is obtained. As $J$ increases, in-gap spectral weight gradually develops, accompanied by a simultaneous suppression of the coherence peaks. When $J$ reaches the strong-coupling regime, such as 2 $t$ to 3 $t$, the gap becomes nearly fully filled and the coherence peaks are largely removed, reproducing the main spectral evolution observed experimentally on the Fe nano-skyrmion lattice.

We further calculated the same DOS ratio used in the experimental analysis, comparing the averaged in-gap DOS with the averaged DOS around the coherence peaks. The ratio increases with $J$ and approaches 1 when the superconducting gap is almost completely filled. Since the chemical potential ($\mu$) and Rashba spin-orbit coupling strength ($\alpha$) can also affect the low-energy DOS, we further calculated the $J$-dependent DOS ratio for different values of these parameters. As shown in Fig. 4(e,f), the ratio shows a similar increase with $J$ for different $\mu$ and $\alpha$, and approaches 1 in the strong-coupling regime. This robust trend indicates that the superconducting-gap quenching is mainly controlled by the onsite exchange coupling $J$, rather than by fine tuning of $\mu$ and $\alpha$. These calculations suggest that the Fe nano-skyrmion lattice on proximitized Ir(111)/Nb(110) lies in a strong exchange-coupling regime, where direct magnetic hybridization with the superconducting substrate becomes strong enough to destroy the local superconducting gap.

**Conclusions**

In conclusion, by growing a thin Ir(111) film on Nb(110), followed by sub-monolayer Fe deposition, we realized a hybrid system of a *field-free* atomic-scale nano-skyrmion lattice and a proximitized superconducting thin film. Low-temperature STS data reveals that superconductivity is completely quenched on the Fe nano-skyrmion lattice, while the surrounding uncovered Ir surface still retains a partially suppressed superconducting gap. Tight-binding BdG calculations reproduce this spectral evolution by increasing the local s-d exchange coupling between the Fe spin texture and the proximitized electronic states, placing Fe/Ir(111)/Nb(110) in a strong exchange-coupling regime. This finding highlights an important but often underappreciated constraint in skyrmion-superconductor platforms. The exchange coupling in such real hybrid systems can be strong enough to destroy the superconducting gap. Future efforts to engineer skyrmion-based topological superconductivity therefore have to take into account considerations of the exchange-coupling strength, such that direct magnet-superconductor coupling can be achieved without quenching the superconducting gap required for topological states.

**Acknowledgement:**

We thank Jens Wiebe, Philip Beck, Thore Posske, and Deliang Bao for helpful discussions. R.W., D.W. and W. L. gratefully acknowledge funding by the EU via the ERC Advanced Grant ADMIRE (No. 786020), the DFG via the Cluster of Excellence “Advanced Imaging of Matter” (EXC 2056, project ID 390715994). D.W. gratefully acknowledges funding by National Natural Science Foundation of China (No. 1247447).

**Author contributions:**

W. L. and R. W. conceived this research. W. L. completed the investigation of the sample growth; D. W. and W. L. performed the low-temperature STM/STS measurements and analyzed the data in discussion with R. W.; E. M. performed the tight-binding calculations; D. W. and W. L. wrote the initial draft of the manuscript and finalized it with input from all authors. All authors contributed to and approved the final manuscript.

**Competing interests:**

The authors declare no competing interests.

**Materials & Correspondence**

Correspondence and requests for materials should be addressed to Dongfei Wang and Wenbin Li.

**Figures:**

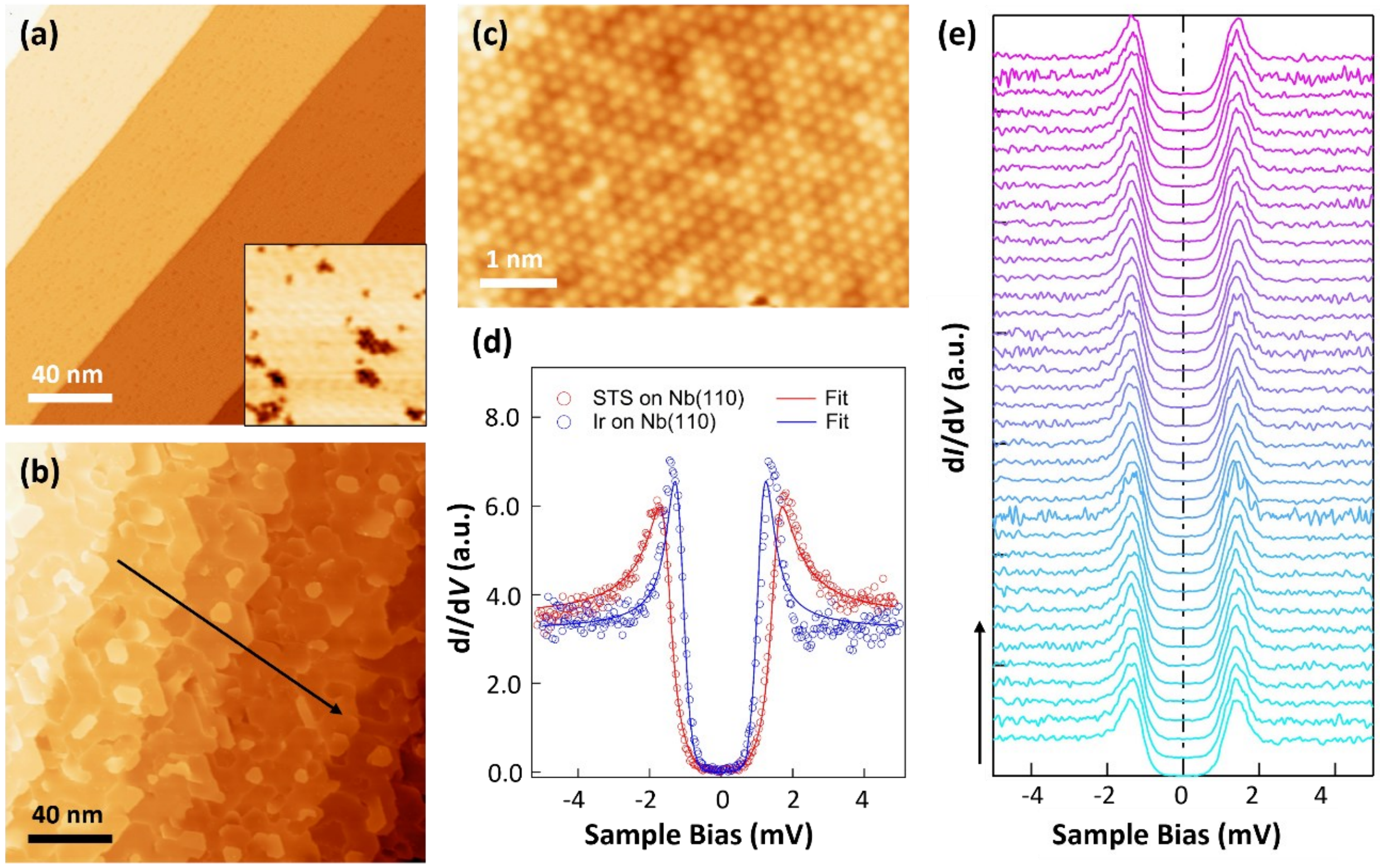

**Fig. 1. STM topography and dI/dV spectra of Ir(111) thin films grown on Nb(110). (a)** STM topography of the clean Nb(110) surface, showing atomically flat terraces separated by step edges. The inset shows a zoomed-in STM image of a 15 nm × 15 nm area, revealing a clean Nb(110) surface with a small amount of dark oxide defects. (**b**) Large-area STM topography of an Ir(111) thin film grown on Nb(110). The Ir film thickness is approximately 3-5 nm. (**c**) Atomically resolved STM image of the Ir(111) thin film, showing the hexagonal lattice structure with a lattice periodicity of about 2.7 Å, corresponding to the Ir(111) surface. (d) Representative dI/dV spectra measured on clean Nb(110) and on the Ir(111)/Nb(110) surface, respectively. Solid lines are fits using the Dynes equation, giving a superconducting gap size of 1.5 meV and 1.2 meV, respectively. (e) Spatially resolved dI/dV spectra acquired along the arrow marked in (b), showing the site-independence of the superconducting gap. STM imaging parameters: (a) V = -1.0 V, I = 2 nA; inset in (a), V = -1.0 V, I = 110 pA; (b) V = -0.1 V, I = 200 pA; (c) V = -30 mV, I = 1.0 nA. STS parameters in (d) and (e): $V_{\text{sample}}$ = -5 mV, I = 800 pA, $V_{\text{mod}}$ = 50 μV, $f_{\text{lock-in}}$ = 1800 Hz.

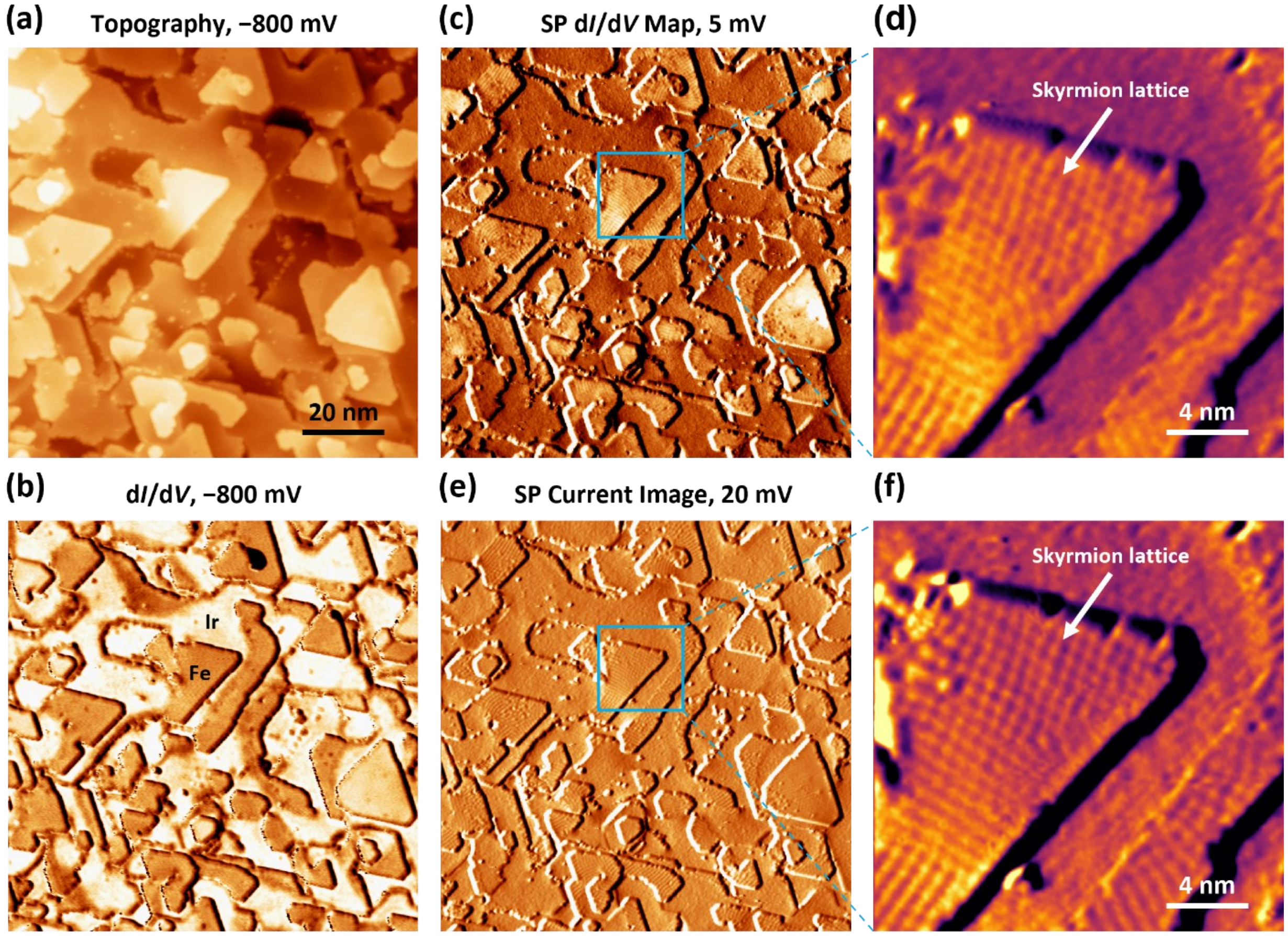

**Fig. 2. STM topography and corresponding dI/dV maps of sub-monolayer Fe on Ir(111)/Nb(110).** (a) STM topography of a representative area after sub-monolayer Fe deposition on the Ir(111) thin film prepared on Nb(110). (b) Corresponding dI/dV map acquired at -800mV, where Fe islands and the uncovered Ir(111) surface can be distinguished by their differential tunneling conductance contrast. (c) Spin-polarized dI/dV map at 5 mV acquired in the same area as in (a), revealing the magnetic contrast of the Fe islands. (d) Zoomed-in spin-polarized dI/dV map of the region marked by the blue rectangle in (c), showing the nanoscale skyrmion lattice on the Fe island. (e) Spin-polarized current map acquired in the same area as in (a), this image includes all the spin polarized tunneling events for energies from 0 to 20 meV. (f) Zoomed-in spin-polarized current map of the marked region in (e), clearly resolving the skyrmion lattice. Measuring parameters: (a,b) V = -800 mV, I = 2 nA, $V_{mod}$ = 50 mV, $f_{lock\text{-}in}$ = 4871.14 Hz; (c, d) $V_{sample}$ = 5 mV, $I$ = 500 pA, $V_{mod}$ = 2 mV, $f_{lock\text{-}in}$ = 4871 Hz. (e, f) $V_{sample}$ = 20 mV, $I$ = 9 nA.

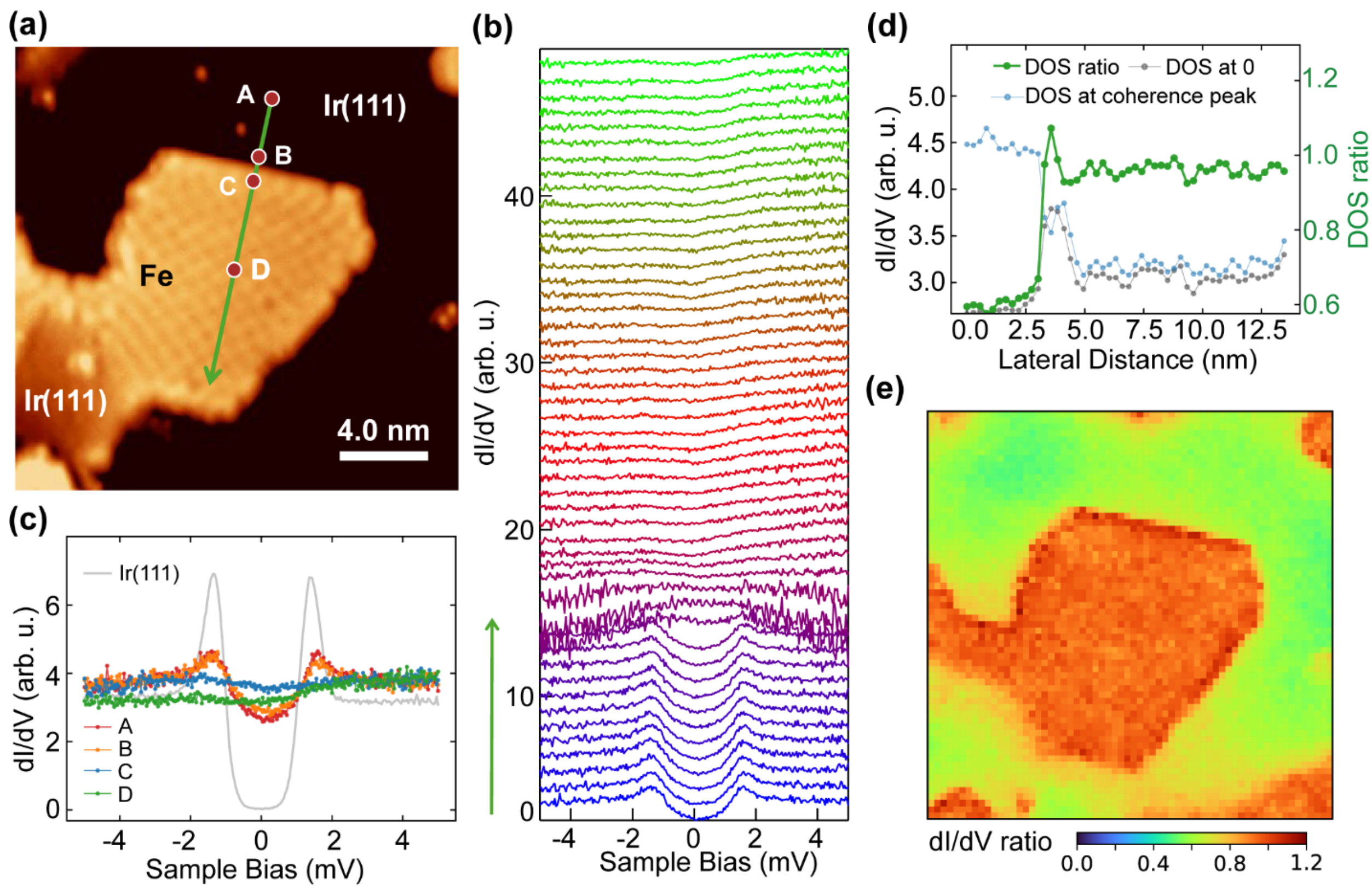

**Fig. 3. Spatially resolved superconducting-gap spectroscopy across an Fe island with a nanoskyrmion lattice on Ir(111)/Nb(110).** (**a**) Spin-polarized STM topography of an Fe island on Ir(111)/Nb(110), where the nano-skyrmion lattice is resolved on the Fe island. (**b**) From bottom to top, spatially resolved dI/dV spectra acquired along the green arrow in (a), showing the evolution of the low-energy spectra from the uncovered Ir(111) surface to the Fe island. (**c**) Representative dI/dV spectra measured at the selected positions A-D in (a), compared with a reference spectrum measured on the clean Ir(111) thin film before Fe deposition. (d) Line profiles of the dI/dV-derived DOS intensities at the Fermi level and the coherence peak, together with their ratio, extracted from the spectra in (b). (e) Map of the dI/dV ratio, defined as dI/dV(0 mV) / {[dI/dV(-1.5 mV) + dI/dV(+1.5 mV)]/2}, corresponding to the same area in (a), visualizing the local suppression of superconductivity around the Fe nanoskyrmion lattice. Measurement parameters: (a) $V_{\text{sample}}$ = 5 mV, I = 300 pA; (b,c,e) $V_{\text{sample}}$ = 5 mV, I = 1.0 nA, $V_{\text{mod}}$ = 30 μV, $f_{\text{lock-in}}$ = 1637 Hz.

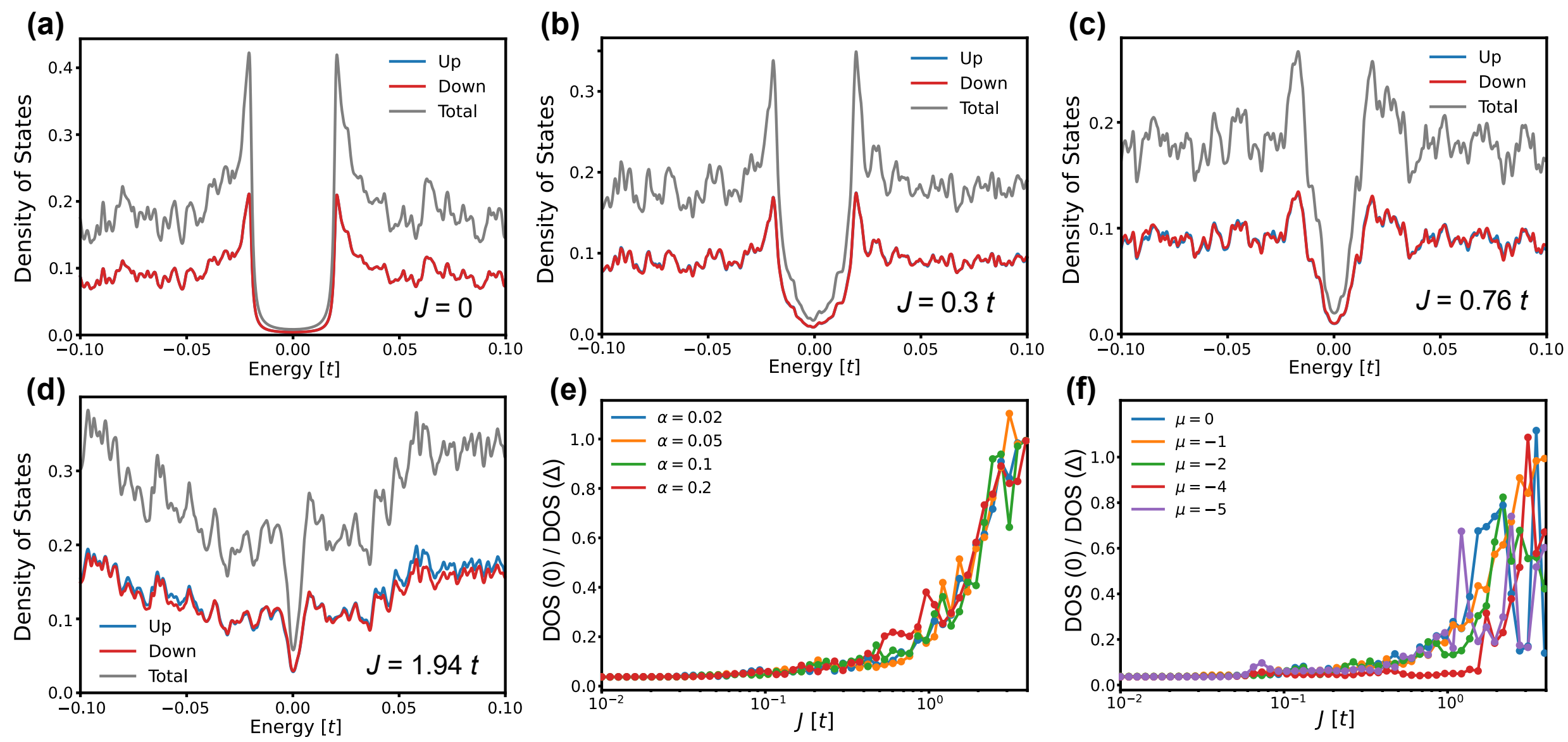

**Fig. 4. Tight-binding calculations for a skyrmion lattice coupled to a superconducting substrate.** (a-d) Calculated density of states near the Fermi energy under different magnet-superconductor coupling strengths $J$, with all other parameters kept fixed. For $J = 0$ in (a), a well-defined superconducting gap with pronounced coherence peaks is obtained. With increasing $J$, the superconducting gap is progressively filled, accompanied by a simultaneous suppression of the coherence peaks. When $J$ reaches approximately 2 $t$ in (d), the superconducting gap is nearly fully quenched. (e,f) $J$-dependent DOS ratio calculated for different spin-orbit coupling strengths α and chemical potentials μ. The DOS ratio is defined as DOS(0) / {[DOS(-Δ) + DOS(+Δ)]/2}, where DOS(0) is the DOS at the Fermi energy and DOS(±Δ) represents the DOS at the superconducting coherence-peak energies. A DOS ratio approaching unity indicates an almost complete filling of the superconducting gap, corresponding to the quenching of superconductivity. Simulation parameters: (a-d) $\mu = -1.0$ $t$, $\alpha = 0.02$ $t$, $\Delta = 0.02$ $t$. (e) $\mu = -1.0$ $t$, $\Delta = 0.02$ $t$. (f) $\alpha = 0.02$ $t$, $\Delta = 0.02$ $t$.